%% file: main.tex
\documentclass[conference]{IEEEtran}
\IEEEoverridecommandlockouts
\usepackage{cite}
\usepackage{amsmath,amssymb,amsfonts}
\usepackage{graphicx}
\usepackage{textcomp}
\usepackage{xcolor}
\def\BibTeX{{\rm B\kern-.05em{\sc i\kern-.025em b}\kern-.08em
    T\kern-.1667em\lower.7ex\hbox{E}\kern-.125emX}}

\include{helper_figures}

\begin{document}

\title{Heterogeneous Ground-Air Autonomous Vehicle Networking in Austere Environments: Practical Implementation of a Mesh Network in the DARPA Subterranean Challenge
\thanks{This work was supported through the DARPA Subterranean Challenge, cooperative agreement number HR0011-18-2-0043, as well as by agreement with Meshmerize GmBH.}
\thanks{$\dagger$ Both authors contributed equally to this work.}
}

\author{\IEEEauthorblockN{Harel Biggie$\dagger$}
\IEEEauthorblockA{\textit{Department of Computer Science} \\
\textit{University of Colorado Boulder}\\
Boulder, USA \\
harel.biggie@colorado.edu}
\and
\IEEEauthorblockN{Steve McGuire$\dagger$}
\IEEEauthorblockA{\textit{Department of Electrical and Computer Engineering} \\
\textit{University of California Santa Cruz}\\
Santa Cruz, USA \\
steve.mcguire@ucsc.edu}

}

\maketitle

\begin{abstract}
Implementing a wireless mesh network in a real-life scenario requires a significant systems engineering effort to turn a network concept into a complete system. This paper presents an evaluation of a fielded system within the DARPA Subterranean (SubT) Challenge Final Event that contributed to a 3rd place finish. Our system included a team of air and ground robots, deployable mesh extender nodes, and a human operator base station. This paper presents a real-world evaluation of a stack optimized for air and ground robotic exploration in a RF-limited environment under practical system design limitations. Our highly customizable solution utilizes a minimum of non-free components with form factor options suited for UAV operations and provides insight into network operations at all levels. We present performance metrics based on our performance in the Final Event of the DARPA Subterranean Challenge, demonstrating the practical successes and limitations of our approach, as well as a set of lessons learned and suggestions for future improvements. 
\end{abstract}

\begin{IEEEkeywords}
mesh networking, ROS, field robotics, system evaluation
\end{IEEEkeywords}


\scaledfig{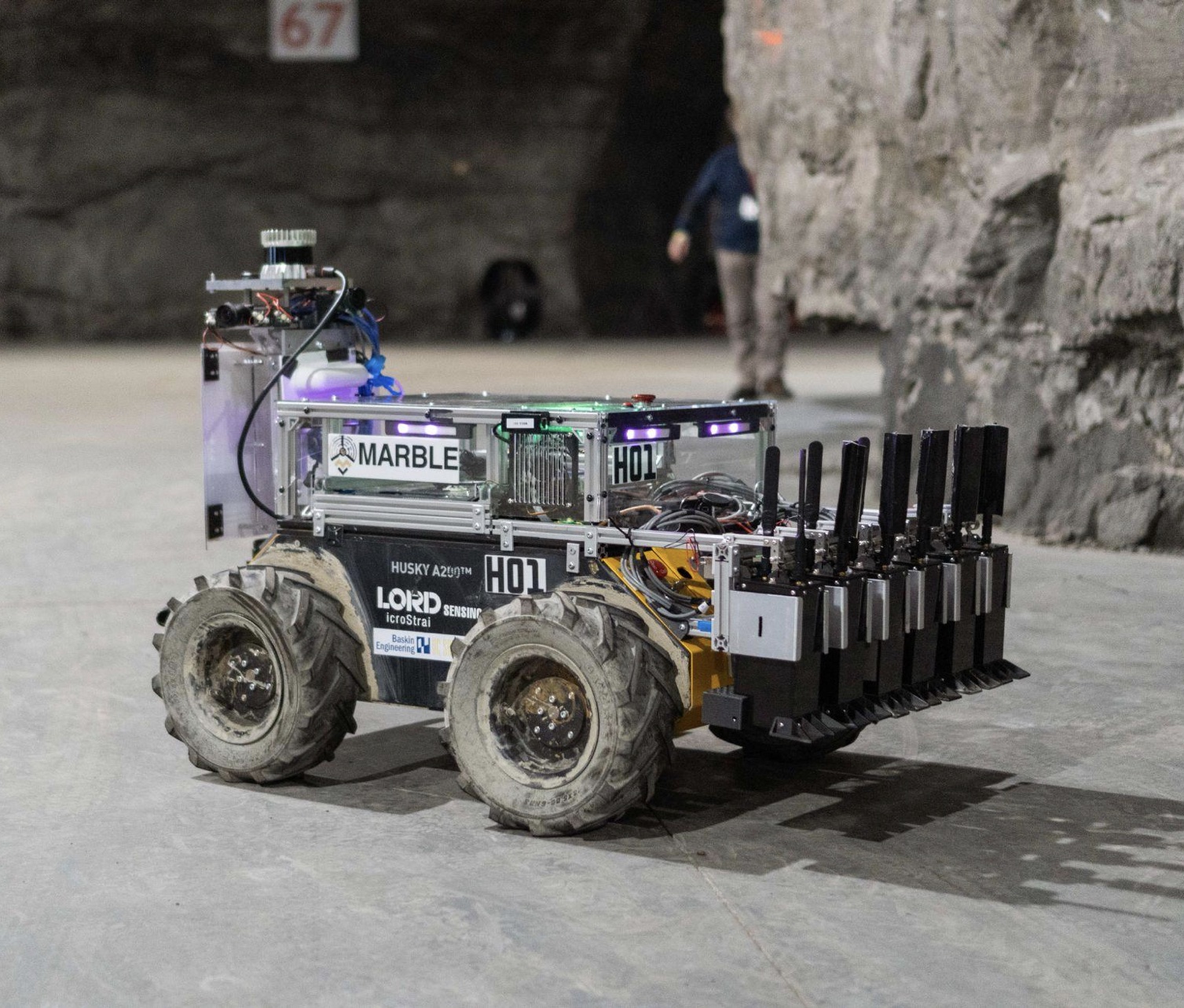}{final_husky}{Husky ground vehicle with 6 comm beacons\vspace{-10pt}}{0.48}

\section{Introduction}
Practical field robotics exploration experiments often require a wireless communications backhaul in austere conditions, where there is no existing infrastructure. In these scenarios, a mesh network, self deployed as part of the exploration task, becomes an ideal solution for extending comm network coverage deep into unknown environments. This paper details a networking solution developed as part of the DARPA Subterranean (SubT) Challenge; a multi-year international effort designed to test the current state of the art in robotic exploration. The Subterranean Challenge consisted of three events in which a single human supervisor directed one or more mobile robotic platforms to map out an unknown environment while searching for DARPA-specified items; the Challenge was designed to simulate a search-and-rescue scenario. The three events were held in a research coal mine, an abandoned nuclear power plant, and a former limestone mine; the use of environments dominated by solid rock and/or concrete presented formidable difficulties for RF communication. This paper presents the development of our network solution, encompassing physical, logical, and transport layers, validated in testing and by our third place finish at the final event; we emphasize our novel contribution, a transport layer called \textit{udp\_mesh}. Additionally, our modular pipeline enables the networking solution to run on a variety of commercially available hardware, in compact form factors making it an ideal choice for UGV and UAV operations alike. Our system has application in any field robotics context (for example, search and rescue or exploration) using mesh infrastructure, with a wide range of implementation options supporting UAV and ground vehicles with equal ease. 

We first present the system requirements as either specified by DARPA or determined through design, then introduce several examples of existing systems that partially fulfill those requirements to motivate the development of our own system. We then describe our complete infrastructure, including several lessons learned through design iterations over the course of the Challenge. Finally, we present a detailed analysis of the performance of our prize-winning system at the final event, with potential directions for future work.

\section{Problem Requirements}

Our system design had to fulfill a multitude of requirements, some of which were specified to our team from DARPA as part of the competition, while others were practical implications dictated by our team composition and infrastructure.

As part of the competition setup, DARPA required any RF system to operate within FCC limitations; while the use of licensed bands was not forbidden, we quickly made the determination that the use of ISM bands reduce administrative burdens needed to test our system in the field. Further, all data products produced by our exploration system had to be funneled through a single wired Ethernet port co-located with a human supervisor monitoring the robots at the base station located outside of the exploration area. Finally, DARPA did not guarantee any kind of spectrum conditions or interference metrics. 

From our team's perspective, required elements of our solution included SWaP-C (size, weight, power, and cost) constraints on hardware, as well as software requirements for integration. Our network infrastructure had to be available in sufficient form factor options to permit deployment on aerial robots and various sizes of ground robot. Further, the cost per node was limited to approximately \$1k, with an emphasis on cheaper hardware. From the software perspective, the communication system had to support a ROS1 (simplified as ROS) \cite{quigley2009ros} environment running on Linux; we needed to support ROS natively to reduce needed integration efforts and enable testing of higher-level functions in simulation. 

From a systems design perspective, we required our solution to have minimal runtime configuration requirements, node discovery, online/offline detection, guaranteed transmission, and traffic prioritization. We desired fast reconnect times, but were willing to trade off with decreased throughput if needed.

\section{Background}
Previous work has developed several solutions to common problems encountered with deploying mesh networks. As details about vendor-specific implementations are often unavailable, we detail several open-source or readily available commercial options that were considered as our system design proceeded.

\subsection{Physical Layer}
Physically, utilizing a standard 802.11 waveform and PHY has significant advantages in that there are many chipset vendors to draw from. We desired a chipset with well-understood Linux support, potentially yielding on the latest features such as high-density MIMO connectivity. Using a standard 802.11 waveform on an unlicensed frequency enables the unregulated use of this system worldwide, potentially also supporting integration with non-mesh networks using layers such as \textit{mac80211} on Linux. As 802.11 is ubiquitous, our system could be used in a wide range of applications and mobile vehicles, including search and rescue, agriculture, and self-deploying mesh networks.


\subsection{Logical Layer}
Meshing layers lay between the physical transmission of frames over the medium and a higher-level protocol such as IP. For applications involving rapid changes to mesh topology, a responsive mesh layer that minimizes lost link time is a major requirement. Further, to reduce integration effort, a mesh layer that operates at layer 2 of an OSI stack\footnote{\url{https://en.wikipedia.org/wiki/OSI_model}} is desirable to allow transparent use of higher-level protocols such as ARP and IP. 
Typically, meshing algorithms such as OLSR \cite{clausen2003optimized} and AODV \cite{perkins2003rfc3561} select a single best path for routing between nodes which hinders algorithmic performance in dynamic environments. A more recent example of a single-path logical meshing layer is Better Approach to Mobile Ad-hoc Networking-Advanced (\textit{batman-adv}) \cite{seither2011routing}, an open source implementation of a layer 2 mesh. 
In contrast to \textit{batman-adv}, \textit{meshmerize} \cite{pandi2019meshmerize} provides multiple paths between nodes to ensure a reliable connection while still operating at layer 2; these multiple paths allow for a dramatic decrease in reconnect times when mesh topology changes. 
Other options for meshing layers such as the original implementation of \textit{batman} operate at layer 3 as packet routers; a downside of this class of mesh layers is a more complicated IP addressing scheme needed over the entire mesh. 

\subsection{Transport Layer}

Many robotic systems utilize middleware layers based on the publisher and subscriber model to coordinate activities of different parts of the system; these middlewares are designed to enable both inter- and intra-compute node connectivity. In general, a middleware layer needs to provide discovery, data encapsulation, and transport services. In many applications, the concept of \textit{prioritization} can also be modeled to reserve access to a limited communication channel and ensure high-priority data (such as an emergency stop request) is given preferred treatment. Contemporary examples include ROS\cite{quigley2009ros}, OROCOS\cite{bruyninckx2001open}, and YARP\cite{paikan2015communication}. While we use ROS as a case study of a publish/subscribe architecture to motivate the requirements of a transport layer, any architecture that utilizes a centralized directory-like service is likely to face similar limitations.

In a traditional networked ROS architecture, a single computer runs a special node known as the \textit{rosmaster} that coordinates the publish-subscribe mechanisms. When a node wishes to exchange data with another node via named \textit{topics}, the master is consulted to determine the computer to connect to, as in \fig{basic_ros}. A single \textit{rosmaster} serves as a central directory of nodes and topics; when a subscription to a topic is requested, a list of publisher nodes is returned so that point-to-point TCP connections can be made directly between publisher and subscriber. When computers are connected via always-on, high-speed links such as wired Ethernet, this system works quite well. However, when computers are connected by an unreliable mesh network, this single-master, TCP-dependent model breaks down; if the master suddenly becomes unavailable, dependent nodes become unreachable. 

\tikzfigcol{basic_ros}{A basic single-master ROS network node graph. Red lines indicate data transfer, where black lines indicate directory management.}{
\tikzstyle{component} = [ text centered, rounded corners,font=\Large, inner sep=5pt, outer sep=0pt, fill=Goldenrod, text opacity=1, text=black, draw, minimum width=2cm]
\begin{tikzpicture}[scale=1, transform shape, post/.style={shorten >=1pt,-{Stealth[length=2mm]},semithick,rounded corners=5pt},semithick]
\node [component,fill=blue!20] (master) {Master};
\path (master) +(3cm, 1cm) node [component] (node_a) {Node A};
\path (master) +(3cm, 0cm) node [component] (node_b) {Node B};
\path (master) +(3cm, -1cm) node [component] (node_c) {Node C};

\draw[post, <->] (node_a.west) -- (master.east); 
\draw[post, <->] (node_b.west) -- (master.east); 
\draw[post, <->] (node_c.west) -- (master.east); 

\draw[post,draw=red, ultra thick] (node_a.east) to[out=-20,in=20,distance=0.5cm] (node_b.east);
\draw[post,draw=red, ultra thick] (node_c.east) to[out=-20,in=20,distance=1cm] (node_a.east);
\end{tikzpicture}
}

To mitigate the risks of a single-master architecture, we used a multi-master architecture in which each robot (and human interface) ran a single master instance (as in \fig{multi_ros}). Within the local network of the robot, the master was always available through a high-speed wired link, resulting in stable local operations. However, additional mechanisms are needed to enable messages to be passed between masters. 

One reference implementation is \textit{multimaster\_fkie} \cite{tiderko2016ros}, which enables nodes connected to different ROS masters to communicate with one another. \textit{multimaster\_fkie} uses broadcasts to discover other instances on the local Ethernet segment; \textit{multimaster\_fkie}\footnote{\url{http://wiki.ros.org/multimaster_fkie}} links independent masters together by cross-publishing their topic and node directories so that TCP connections can be made across \textit{rosmaster} boundaries. Functionally, when a local node wishes to subscribe to a remote master's topic, the local master now contains the remote connectivity information; the node then establishes a standard TCP link between nodes and messages can flow. 

While \textit{multimaster\_fkie} solves the discovery and advertisement problems, it does nothing to establish prioritization of data flow. With the standard TCP transport provided by ROS, there is no centralized means of monitoring inter-node connections to arbitrate data priorities. We are aware of at least two alternatives that provide this critical quality-of-service prioritization within a ROS environment by channelling all inter-robot traffic through a single monitoring point to enforce data priorities.

The first, \textit{Pound}\footnote{\url{https://github.com/dantard/unizar-pound-ros-pkg}} \cite{tardioli2019pound}, is specifically designed for use in unreliable mesh networks and implements many of the desired requirements. However, \textit{Pound} relies on hardcoded topic names and fixed addressing information, which were critical requirements for our system for ease of testing and integration.

Alternatively, \textit{nimbro\_network} \cite{schwarz2016supervised} implements a similar set of functions with regards to transport over wireless networks, but omits prioritization. Crucially, \textit{nimbro\_network} still utilizes TCP for reliable inter-robot communication, preventing adaptation of core TCP behavior (particularly retransmits) to unreliable mesh networks; UDP links are only used for non-guaranteed data delivery.

We sought to design a transport layer for use in a ROS environment that allowed for runtime reconfiguration, implemented prioritization, and re-implemented reliable communication over UDP to have better control over retransmits and fragmentation. 

\tikzfigcol{multi_ros}{A multi-master ROS network node graph. Red lines indicate data transfer, where black lines indicate directory management. Blue lines are data paths that cross network segments.}{
\tikzstyle{component} = [ text centered, rounded corners,font=\Large, inner sep=5pt, outer sep=0pt, fill=Goldenrod, text opacity=1, text=black, draw, minimum width=2cm]
\def\masterdist{1cm}
\begin{tikzpicture}[scale=1, transform shape, post/.style={shorten >=1pt,-{Stealth[length=2mm]},semithick,rounded corners=5pt},semithick]
\node [component,fill=blue!20] (master) {Master $\alpha$};
\path (master) +(0cm, -1cm) node [component] (node_a) {Node A};
\path (master) +(0cm, -2cm) node [component] (node_b) {Node B};
\path (master) +(0cm, -3cm) node [component] (node_c) {Node C};

\path (master) +(5cm, 0cm) node [component,fill=blue!20] (master_b) {Master $\beta$};
\path (master_b) +(0cm, -1cm) node [component] (node_d) {Node D};
\path (master_b) +(0cm, -2cm) node [component] (node_e) {Node E};
\path (master_b) +(0cm, -3cm) node [component] (node_f) {Node F};

\draw[post, <->] (node_a.west) to[out=200,in=160,distance=0.75cm] (master.west); 
\draw[post, <->] (node_b.west) to[out=200,in=160,distance=0.75cm] (master.west); 
\draw[post, <->] (node_c.west) to[out=200,in=160,distance=0.75cm] (master.west); 

\draw[post, <->] (node_d.east) to[out=20,in=0,distance=0.75cm] (master_b.east); 
\draw[post, <->] (node_e.east) to[out=20,in=0,distance=0.75cm] (master_b.east); 
\draw[post, <->] (node_f.east) to[out=20,in=0,distance=0.75cm] (master_b.east);

\draw[post,draw=red, ultra thick] (node_a.east) to[out=-20,in=20,distance=0.5cm] (node_b.east);
\draw[post,draw=red, ultra thick] (node_c.east) to[out=20,in=20,distance=1cm] (node_a.east);

\draw[post,draw=red, ultra thick] (node_d.west) to[out=200,in=160,distance=0.5cm] (node_e.west);
\draw[post,draw=red, ultra thick] (node_f.west) to[out=160,in=160,distance=1cm] (node_d.west);

\draw[post,draw=blue, ultra thick] (node_b.east) to[out=-20,in=200,distance=0.5cm] (node_e.west);
\draw[post,draw=blue, ultra thick] (node_f.west) to[out=200,in=-20,distance=0.5cm] (node_c.east);
\path (master.north)+(2.5cm,0cm) node (mesh_hi) {};
\path (node_c.south)+(2.5cm,0cm) node (mesh_lo) {};
\draw[dotted, ultra thick,gray] (mesh_hi) -- (mesh_lo);

\end{tikzpicture}
}



\section{Methods}

\subsection{Physical}
 We utilized \textit{ath9k}-supported commercially available 802.11 hardware based on Atheros (and subsequently Qualcomm) chipsets capable of 2x2 multiple-input, multiple-output (MIMO) transmission rates across our entire robot fleet, base stations, and comm beacons. Due to the widespread industry availability of \textit{ath9k} supported hardware using OpenWRT\footnote{\url{https://openwrt.org/}}, we had the freedom to use high-power 2W radios in our deployable comm beacons, 1W radios in our robots, and very low power radios for testing and validation exercises without having to change any underlying code or settings.  Newer chipsets supporting 3x3 MIMO were evaluated, but found to not have the firmware stability in the `ad-hoc' mode used by higher mesh networking layers. Each of our Husky ground robots (\fig{final_husky}) utilized a pair of radios, each with a 2x2 sector antenna pointing fore and aft, while our Spot ground robots (\fig{final_spot}) utilized a pair of omnidirectional antennas on a single radio.

\subsection{Logical}
We partnered with Meshmerize GmbH\footnote{\url{https://www.meshmerize.net/}} to improve upon the \textit{batman-adv} layer 2 open-source mesh networking solution, as well as provide subject-matter expertise.
In our preliminary evaluations, we determined that on average \textit{batman-adv} was able to re-establish connection within 10s while \textit{meshmerize} was able to do so in 1s by prioritizing connectivity over optimal routing.
This fundamentally different architecture enables a data ferrying paradigm; robots need only have brief windows of connectivity to have a functional exploration and reporting strategy. However, \textit{meshmerize}’s throughput is limited to approx 10Mbit/s due to CPU performance limitations on radio system-on-chip processors.
\textit{meshmerize} is capable of running on all \textit{ath9k}-supported hardware using an OpenWRT integration. With \textit{meshmerize} functioning at layer 2 bridged to each robot's internal network, all IP protocols were enabled; being able to transparently use SSH without any additional software was a significant aid in debugging and troubleshooting.

\tikzfigcol{robot_network}{On-robot network architecture for the Husky platform. Dual mesh radios are bridged to provide mesh access for onboard compute and motion control. Spot and UAV platforms similar.}{
\tikzstyle{component} = [ text centered, rounded corners,font=\Large, inner sep=5pt, outer sep=0pt, fill=green!20, text opacity=1, text=black, draw, minimum width=2cm]
\def\masterdist{1cm}
\begin{tikzpicture}[scale=1, transform shape, post/.style={shorten >=1pt,-{Stealth[length=2mm]},semithick,rounded corners=5pt},semithick]

\node [component,fill=blue!20] (radio) {Mesh radio};
\path(radio) +(-2cm,0) node[antenna] (ant_a) {};

\path (radio) +(3cm,0) node [component,fill=blue!20] (radio_b) {Mesh radio};
\path(radio_b) +(2cm,0) node[antenna] (ant_b) {};

\path (radio) +(1.5cm, 1cm) node [component] (switch) {Ethernet Switch};
\path (switch) +(0cm, 1cm) node [component, rotate=90, anchor=west] (cpu) {CPU};
\path (switch) +(-1cm, 1cm) node [component, rotate=90, anchor=west] (basectl) {Platform};
\path (switch) +(1.5cm, 1cm) node [component, rotate=90, anchor=west, fill=red!20] (lidar) {Lidar};

\draw[post, <->] (lidar.north) -- (cpu.south);
\draw[post, <->] (cpu.west) -- (switch.north);
\draw[post, <->] (switch.161) -- (basectl.west);
\draw[post, <->] (switch.200) -- (radio.70);
\draw[post, <->] (switch.340) -- (radio_b.110);

\draw[post] (ant_a) -- (radio.west);
\draw[post] (ant_b) -- (radio_b.east);


\end{tikzpicture}
}

\subsection{Transport}
The main innovation in our system is our transport layer, \textit{udp\_mesh}. We designed \textit{udp\_mesh} to replace \textit{fkie\_multimaster} to address the need to be able to prioritize message transmissions and provide for better performance over intermittently connected networks. As implemented, our ROS interconnect layer now provides the following services: discovery, address resolution, ROS message encapsulation, point-to-point transport, point-to-multipoint transport, and quality-of-service prioritization.

Each of these services is utilized to support the higher-level communications functions needed for mission success. Fundamentally, the \textit{udp\_mesh} layer uses only unicast and broadcast UDP datagrams to implement higher-level services without requiring multicast support. In principle, multicasting would offer a performance benefit by reducing broadcast traffic. However, in a wireless mesh environment, these potential gains are offset by multicast group membership management overhead.

\subsubsection{Discovery and Address Resolution}
\textit{Discovery} in our layer is the process of identifying nodes that are available for communication. We implement discovery through the use of a periodic heartbeat broadcast that advertises the node's availability and provides name resolution information. In concept, this service is similar to the \textit{mcast\_dns} service in Linux, where peers advertise their naming information to be able to address nodes by hostname instead of layer 2 MAC or layer 3 IP address. Nodes identified through discovery are added to the list of available nodes for communication as well as status reporting. This discovery heartbeat is also used as a lost-communications detector to prevent higher-level messages from queueing for unreachable nodes. 

\subsubsection{ROS Message Encapsulation}
In the ROS ecosystem, messages are translated from a message definition language specification into internal representations appropriate to the implementing language\footnote{\url{http://wiki.ros.org/msg}}. This same language specification is used to serialize and deserialize messages; that is, to transform a ROS message into a buffer of bytes suitable for transmission over an arbitrary channel. \textit{udp\_mesh} implements a generic message passing system such that the message to be transmitted is never deserialized, saving a significant amount of processing time in the case of complex, large message types such as images. Instead, a generic subscriber is used to acquire the serialized bytes for direct use to be transmitted to other nodes. On the receiver side, the transmitted byte stream is deserialized to instantiate the message in a format that other ROS ecosystem nodes can readily consume. These two functions abstract the transport of arbitrary messages over the \textit{udp\_mesh} layer and remove any requirement to define a list of acceptable message types.


\subsubsection{Point to Point Transport}
In the \textit{udp\_mesh} system, point-to-point transport is implemented via UDP datagrams. This envelope contains provisions for sequence tracking, fragmentation, and message reconstruction. As part of preparing a message for transmission, the byte buffer provided by the ROS encapsulation service is split into chunks that fit with in the underlying medium's maximum transmit unit (MTU). 
For the standard 802.11 framing that is used in our system, this MTU is 1500 bytes, out of which 100 bytes are reserved for overhead, leaving 1400 bytes for payload out of every datagram.

In the implementation of our system, a configurable number of message fragments are permitted to be `in flight' at any given time, similar to TCP congestion window control. In order for the next fragment to be transmitted, the receiver must send an acknowledgment. During unit testing to determine an appropriate value for the number of in-flight fragments permitted, an initial increase yields improved throughput. However, after a certain point, throughput decreases as multiple packets are queued for transmission on the medium and start to destructively interfere. As a compromise, three packets are permitted to be in-flight between any two nodes at a time. With this configuration, our transport-layer throughput is approximately 
20 Mbit/s of payload data, measured using raw images as representative high-density traffic over a wired gigabit Ethernet link.

Retransmits are automatically queued until either an acknowledgment is received or the host is marked offline due to non-reception of any heartbeat or acknowledgment messages. Once a host is marked offline, any attempts to send messages are discarded. Hosts may become online once again after receipt of a discovery message. On the receiver side, the message is kept in a temporary state while the fragments arrive. Should message fragments stop arriving, the partial message is purged after a timeout and the host is once again marked offline which indicates to higher levels that reliable transport is unavailable. 

\subsubsection{Quality of Service}
\textit{Quality of Service} (QoS) is the notion that some traffic should be prioritized over other traffic for use of a limited communications channel. As observed in prior competition events, sending large chunks of data such as full maps prevented other, more high priority data from being transmitted such as artifact reports and telemetry. Fundamentally, TCPROS (the default transport used in ROS v1) is not capable of implementing a QoS scheme where a limited channel is shared between different topics (\fig{comm_qos}), as every node subscribing to a topic uses an individual TCP point-to-point link with no information about other links. This need to prioritize traffic was the driving rationale behind the development of the \textit{udp\_mesh} layer. As part of the configuration of the layer, each topic to be transported includes a priority number; internally, this priority number is used as a sorting key to order message fragments for transmission.
  
\tikzfigcol{comm_qos}{In contrast to basic multi-master ROS (\fig{multi_ros}), \textit{udp\_mesh} creates a single virtual channel between nodes, shown in green, to implement data prioritization.}{
\tikzstyle{component} = [ text centered, rounded corners,font=\Large, inner sep=5pt, outer sep=0pt, fill=Goldenrod, text opacity=1, text=black, draw, minimum width=2cm]
\def\masterdist{1cm}
\begin{tikzpicture}[scale=0.93, transform shape, post/.style={shorten >=1pt,-{Stealth[length=2mm]},semithick,rounded corners=5pt},semithick]
\node [component,fill=blue!20] (master) {Master $\alpha$};
\path (master) +(0cm, -1cm) node [component] (node_a) {Node A};
\path (master) +(0cm, -2cm) node [component] (node_b) {Node B};
\path (master) +(0cm, -3cm) node [component] (node_c) {Node C};

\path (node_b.east) + (1cm, 0cm) node [component,rotate=90,fill=green!20] (udpmesh_a) {\textit{udp\_mesh}};

\path (master) +(6cm, 0cm) node [component,fill=blue!20] (master_b) {Master $\beta$};
\path (master_b) +(0cm, -1cm) node [component] (node_d) {Node D};
\path (master_b) +(0cm, -2cm) node [component] (node_e) {Node E};
\path (master_b) +(0cm, -3cm) node [component] (node_f) {Node F};

\path (node_e.west) + (-1cm, 0cm) node [component,rotate=90,fill=green!20] (udpmesh_b) {\textit{udp\_mesh}};

\draw[post, <->] (node_a.west) to[out=200,in=160,distance=0.75cm] (master.west); 
\draw[post, <->] (node_b.west) to[out=200,in=160,distance=0.75cm] (master.west); 
\draw[post, <->] (node_c.west) to[out=200,in=160,distance=0.75cm] (master.west); 
\draw[post, <->] (udpmesh_a.east) to[out=90,in=0,distance=0.5cm] (master.east);

\draw[post, <->] (node_d.east) to[out=20,in=0,distance=0.75cm] (master_b.east); 
\draw[post, <->] (node_e.east) to[out=20,in=0,distance=0.75cm] (master_b.east); 
\draw[post, <->] (node_f.east) to[out=20,in=0,distance=0.75cm] (master_b.east); 
\draw[post, <->] (udpmesh_b.east) to[out=90,in=180,distance=0.5cm] (master_b.west);

\draw[post,draw=red, ultra thick] (node_a.east) to[out=-20,in=20,distance=0.5cm] (node_b.east);
\draw[post,draw=red, ultra thick] (node_c.east) to[out=20,in=20,distance=0.75cm] (node_a.east);

\draw[post,draw=red, ultra thick] (node_d.west) to[out=200,in=160,distance=0.5cm] (node_e.west);
\draw[post,draw=red, ultra thick] (node_f.west) to[out=160,in=160,distance=0.75cm] (node_d.west);

\draw[post,draw=blue, ultra thick] (node_b.east) -- (udpmesh_a.north);
\draw[post,draw=blue, ultra thick] (udpmesh_a.north |- node_c.east) -- (node_c.east);
\draw[post,draw=blue, ultra thick] (node_f.west) -- (udpmesh_b.south |- node_f.west);
\draw[post,draw=blue, ultra thick] (udpmesh_b.south) -- (node_e.west);

\draw[post,draw=JungleGreen, ultra thick, <->] (udpmesh_a.south) -- (udpmesh_b.north);

\path (master.north)+(3cm,0cm) node (mesh_hi) {};
\path (node_c.south)+(3cm,0cm) node (mesh_lo) {};
\draw[dotted, ultra thick,gray] (mesh_hi) -- (mesh_lo);

\end{tikzpicture}
}

\subsubsection{Point-to-Multipoint Transport}
Although \textit{udp\_mesh} is based around point-to-point message transfer, mission requirements sometimes necessitate system-wide messaging. To facilitate these type of messages, a broadcast mechanism is provided by the transport layer. For messages that fit within a single MTU, a single, unacknowledged UDP broadcast is used to distribute the message. For larger messages, individual links to each node are used to send the broadcast as a series of unicast fragments using the same accounting and acknowledgments as the point-to-point mechanism. In both cases, the receiver is unaware that the message was sent as a broadcast versus a single, directed message. 

\section{Relevance to UAV Operations}
Our developed mesh networking solution is particularly useful for robotic systems containing unmanned aerial vehicles (UAVs) in difficult RF environments. The scenarios constructed by DARPA specifically limit RF transmission to line-of-sight only, requiring some sort of mesh relay system to maintain communications. With our system, the radios used in our relay nodes and those used in our robots are functionally identical. Using the same \textit{ath9k}-supported chipsets in varying form factors, the radios used exactly the same meshing software, which allowed robots to relay through one another as they roamed about the environment. To leverage these connections of opportunity, the speedy reconnect times of our system are critical. Finally, our system provides an easy way to integrate into the large body of existing work for UAV flight operations built on top of ROS with no changes required. While we were unable to deploy a UAV platform in the final competition, we performed limited tests with a QAV500 UAV utilizing our mesh networking system to smoothly interface with the rest of our robot fleet.


\section{Functional Test}

Our architecture was validated in a testing campaign leading up to our final competition performance. We present an analysis of our prize-winning run to describe several objective measurements, as well as present a useful case-study balancing network loading versus mission objectives. 

For our final run, we deployed four robots (each acting as a dynamic mesh node) to explore the environment. Two of the robots, the Husky ground vehicles (\fig{final_husky}), were equipped with six comm beacons and a dispenser, where each beacon contained a 2W radio running the \textit{meshmerize} stack. Operationally, the Spot quadrupeds (\fig{final_spot}) were deployed first, while the Huskies followed to establish a comm link back to the base station. Nodes were deployed autonomously by higher-level software. \fig{commmap} overlays the generated map (shown in white) with each robot trajectory; the green-blue colors show where each robot was within communication, while the red-magenta colors show a disconnected state. This figure shows that our network was able to maintain communication through a majority of the course. Our base station operator was able to observe live telemetry from each robot, including position and mapping information, while reserving the ability to take command as needed. Over the entire one-hour run, our robots transmitted 125.2MB of data, which included maps, telemetry, and discovered object reports. Our robots also shared data with each other to implement multi-agent coordination, which this total figure does not include. While our robot cohort also included UAV platforms in testing (\fig{final_uav}), no UAVs were deployed in the final competition run due to strategic reasons unrelated to network performance.

\scaledfig{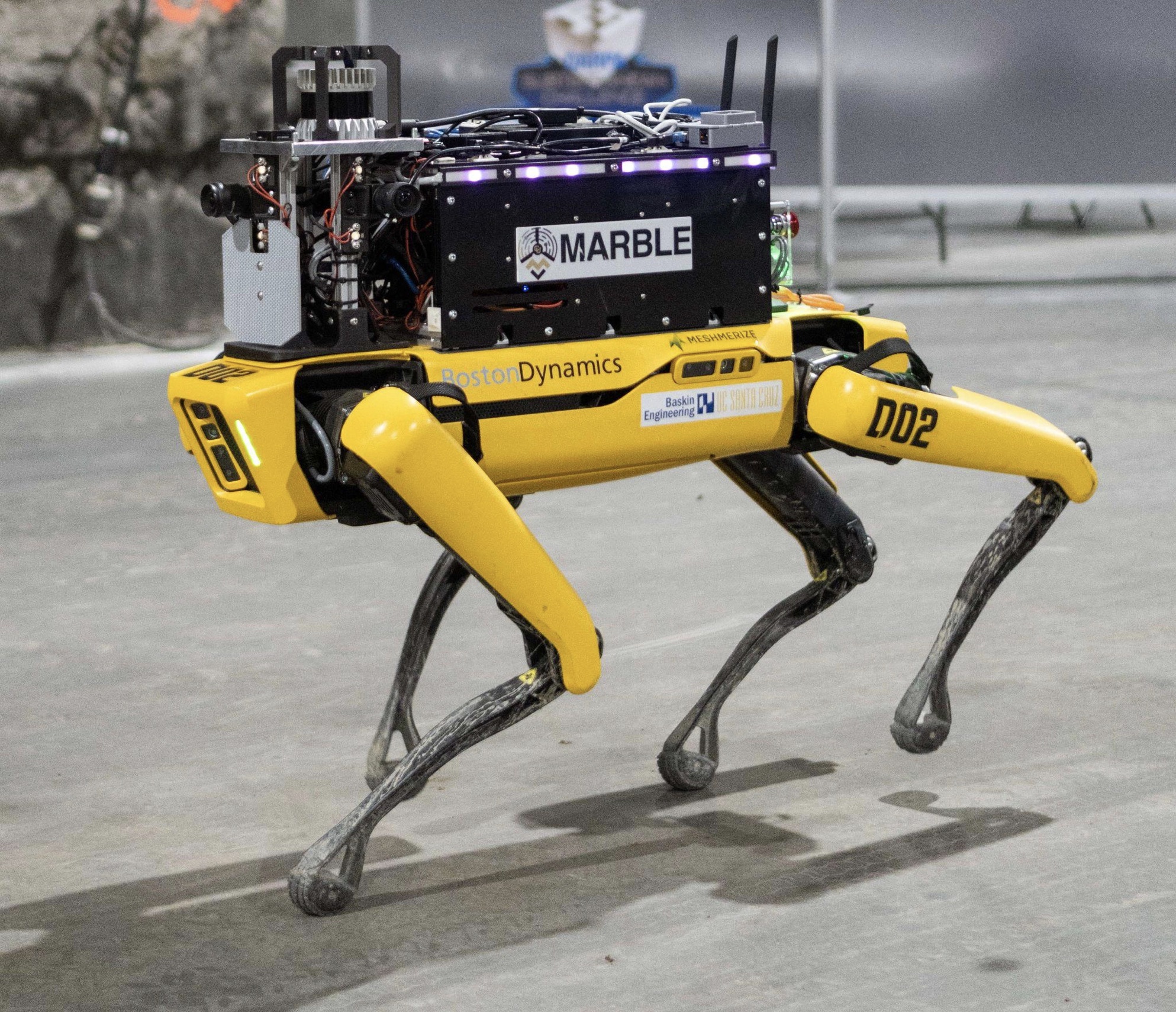}{final_spot}{Spot ground vehicle}{0.48}
\scaledfig{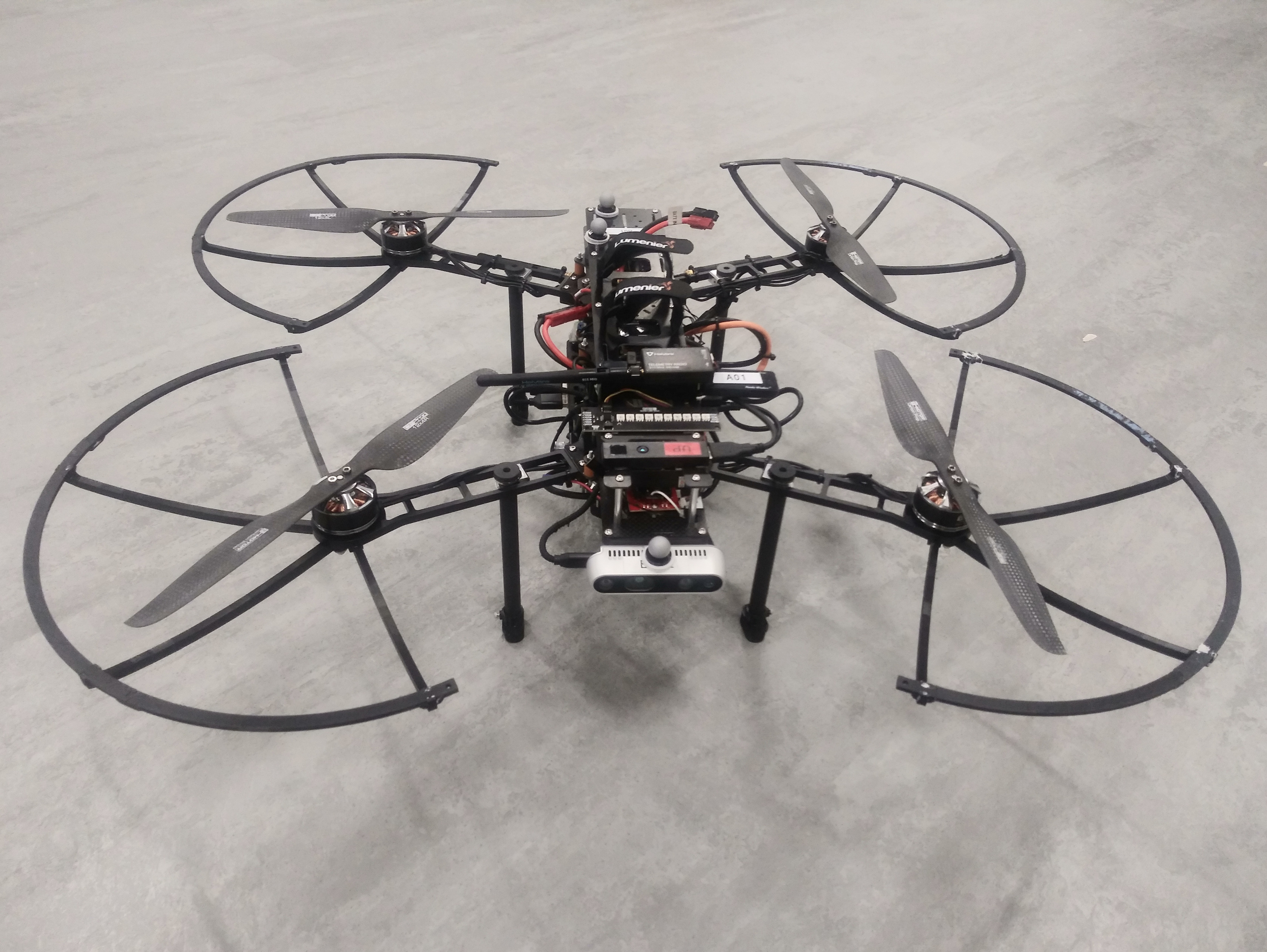}{final_uav}{QAV500 UAV platform designed, but not deployed.}{0.48}

\fig{message_times} plots a PDF of intermessage arrival times for our main management message, as sent from a single robot (D01) to the base station over the duration of our final run. These messages publish at 1 Hz; an ideal system would observe all intermessage arrival times to be 1 second in duration. Due to network latency and usage, we observe an approximately normal ($\mathcal{N}(1.00,0.04)$) distribution of arrival times. Since messages may experience delays, the immediately following message may exhibit an intermessage time of less than one second. Our key observation of this plot is that the bulk of messages arrive within five percent of their expected times across a distance of hundreds of meters and multiple mesh hops. 


\tikzfigcol{message_times}{PDF of intermessage arrival times for D01 with a notional publishing rate of 1 Hz, overlaid with $\mathcal{N}(1.00,0.04)$. The highlighted red region represents the first $\sigma$ value which contains 68\% of the message times.}{\input{figures/message_times.tex}}

\tikzfig{commmap}{htb}{Map from the final run of the DARPA SubT challenge depicting the connection status of each of the four robots as they traversed through the course. Blue-green indicates connected, while red-magenta indicates disconnected. The red star indicates the location of the human operator, base station, and root communication node.}{
\begin{tikzpicture}[scale=0.99, transform shape]
\node (comm_map) {\includegraphics[width=\textwidth,trim=0cm 0cm 0.1cm 0cm,clip]{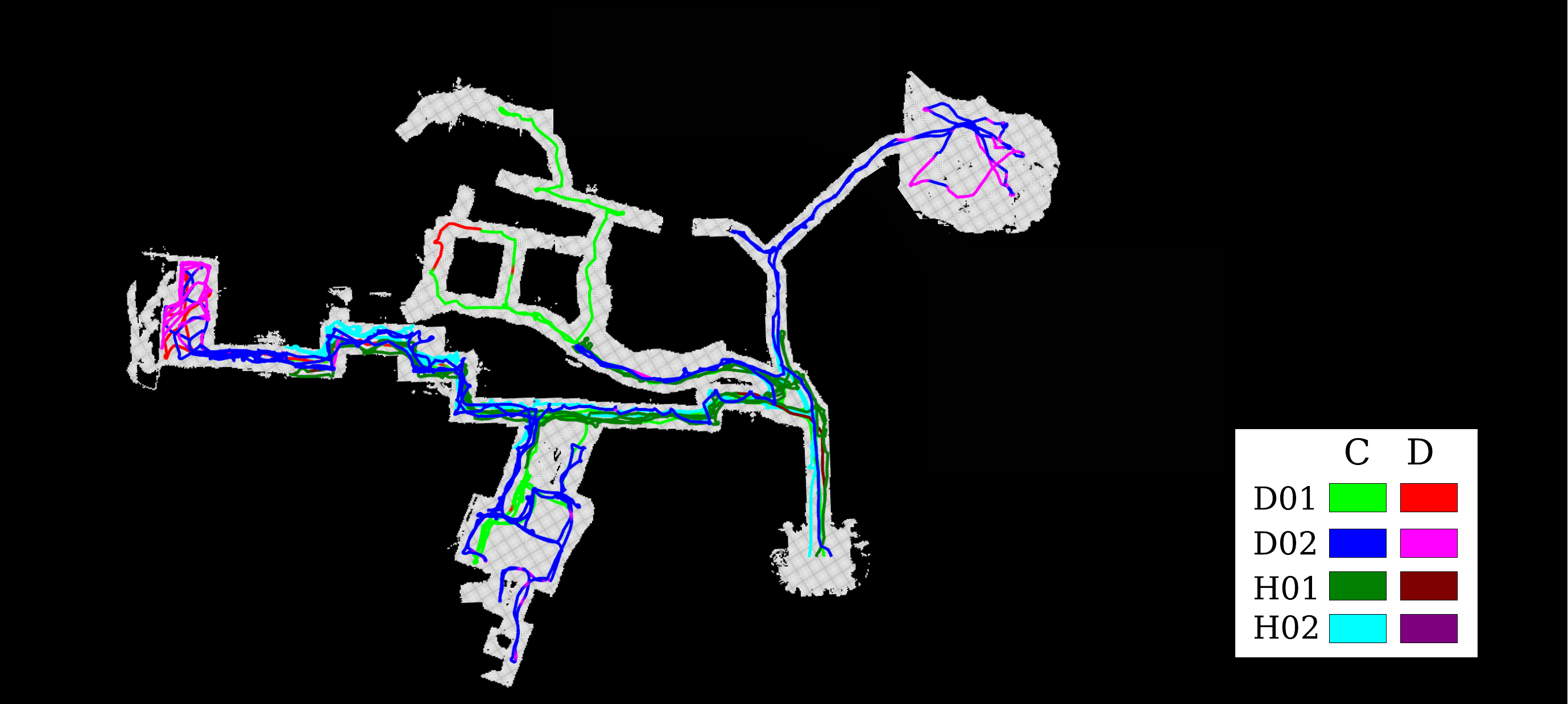}};
\path (comm_map.south) +(0.4cm,1.4cm) node[draw,fill=red,star,star point height=.7em,minimum size=1em]{};
\end{tikzpicture}
\vspace{5pt}
}

\subsection{Case Study: Differential Maps}
\label{diff_maps}
\scaledfig{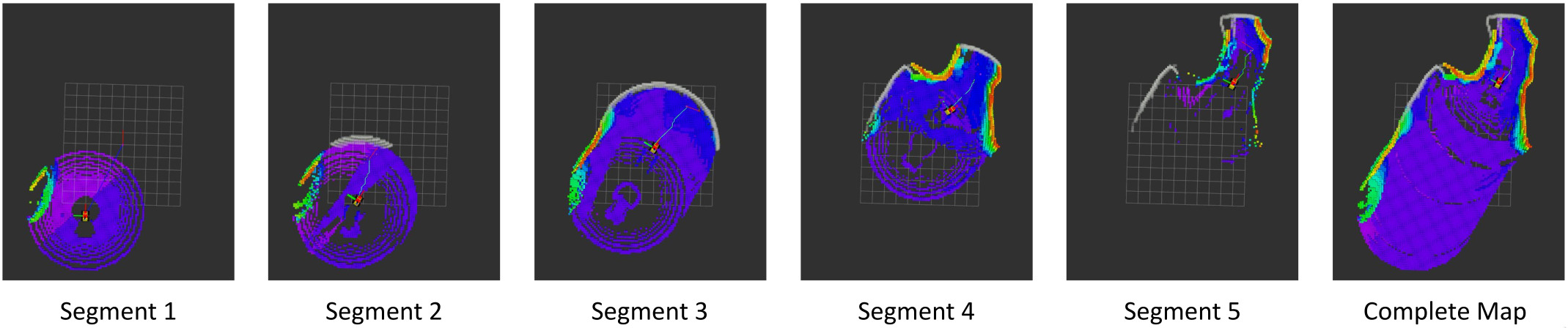}{comm_diff_maps}{Differential map transmission. Instead of a complete map message, incremental map portions are shared between robots, reducing required network bandwidth.}{0.48}
One particular effort from our team to reduce bandwidth requirements was the `differential maps' mechanism used to transmit map portions amongst robots and the base station. Shown in \fig{comm_diff_maps}, each of the five component segments are combined to form the complete map. During a robot's exploration phase, the telemetry data is used to trigger the creation of a new segment to keep individual `diff' elements sufficiently compact. Within our prioritization framework, these diff elements are at a lower priority than artifacts, telemetry, and human supervisor commands. In practice, this innovation helped us to make sure that our limited bandwidth was used for the most important information. As a robot proceeds along its mission, the human supervisor would often have a live view of telemetry and instant command response, while mapping data would gradually fill in as communication bandwidth became available. 

\subsection{Case Study: Adding Point-Of-View Imagery}
\scaledfig{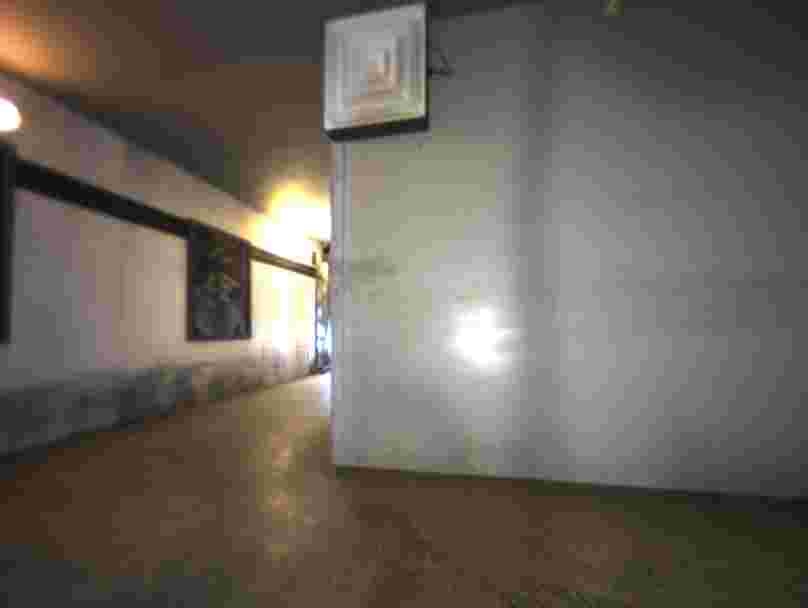}{comm_fpv}{First-person video frame showing compression artifacts and a vent artifact}{0.48}
During our testing campaign leading to the final event, our improved communication stack based on \textit{meshmerize} and \textit{udp\_mesh} had proven to be quite reliable and robust, particularly with respect to link saturation and data prioritization. This confidence in our system allowed our team to be more aggressive in our communications risk posture to utilize our bandwidth to improve scoring. In testing, we enabled point-of-view (POV) video using the front camera from each robot to send highly compressed imagery back to the human supervisor for review. This modification required a single line configuration change, to instruct the \textit{udp\_mesh} layer to transmit the FPV video topic at a priority level below that of all other traffic (telemetry, artifacts, and map data). Sent at 1 Hz and JPEG quality level 10 (out of a maximum of 100), these grainy images with visible compression artifacts (\fig{comm_fpv}) were key to our human supervisor being able to navigate through fog and identify artifacts that our autonomous artifact detection system missed. As a result of our pilot run, this change was rolled out to all robots in the fleet and significantly contributed to our success. In post-run analysis, this live imagery consisted of 51.6MB of the 125.2MB total data received by the base station over our mesh network.

\section{Discussion}
In the development of our system, several key ideas emerged that warrant discussion and analysis. 

\subsection{Porting to arbitrary publish / subscribe architectures}
While we have used ROS as an integration middleware, there is no fundamental tie to the ROS ecosystem that precludes adapting \textit{udp\_mesh} to an arbitrary publish/subscribe architecture. Internally, \textit{udp\_mesh} includes a pair of reader/writer adapters that marshal data to and from the underlying architecture. These adapters could be replaced for any other type of architecture, such as OROCOS \cite{bruyninckx2001open}, which uses Common Object Request Broker Architecture, or Yet Another Robot Platform (YARP) \cite{paikan2015communication}, an architecture specifically designed for quality-of-service management. 

\subsection{Scaling to larger team sizes}
Mesh network systems typically implement broadcast datagrams as a `flood', which requires each node to repeat the message to ensure maximum distribution. Due to the difficulty in managing multicast group membership, broadcast is often utilized for multicast traffic as well. Because \textit{udp\_mesh} relies on broadcast messages to implement discovery, online detection, and name resolution, there is the potential for negative consequences as node count increases. As nodes are added to the network, these broadcast messages result in geometric growth of discovery / advertising traffic. One potential mitigation for this effect relies on eliminating redundant broadcast traffic when a stable link exists between two nodes and messages are being exchanged reliably. Alternatively, \textit{udp\_mesh} nodes could be configured with fixed neighbors to eliminate the broadcast mode.

\subsection{Commodity Hardware}
By relying only on commodity hardware, ranging in price from approximately \$30 to \$1000 each, our team had the freedom to select the hardware realization that best fit our particular needs for each robot class and our beacons without being locked into a particular manufacturer. \textit{ath9k} support has been well-established within the open-source community, with many other users around the world relying on the same hardware. Even though our radios were manufactured by three separate companies, we experienced zero problems related to vendor interoperability. Our development team was physically separated; being able to develop mesh capabilities using budget hardware allowed each team member to have a set of development nodes. 

\subsection{Interoperability with Existing Systems}
Building off of the concept of the well-known OSI network model, we specifically engineered our system so that each layer need not be aware of how other levels are implemented. This modularity is present in the \textit{meshmerize} software being able to support arbitrary IP packets, as well as our \textit{udp\_mesh} being completely ROS message agnostic. As a result, we were able to swap out components on a layer-wise basis with minimal reconfiguration of our systems. Further, by utilizing the ROS ecosystem, our communications system was completely compatible with every ROS node and tool that our team had already written, requiring no code modifications. However, this agnostic behavior meant that there was no awareness of network conditions at higher levels within our autonomy stack. For example, if there were to be a tighter coupling between mesh conditions and autonomy, a robot might adjust its data products to the underlying network conditions (such as improving first-person video quality, etc). Our initial evaluations led to a very conservative estimate of mesh performance; the limited data rate observed at the final event suggests that additional bandwidth was available, but unused.

\subsection{Fast Reconnect Times}
As expected, in early testing underground our radios were not able to penetrate rock formations and concrete which limited us to line of sight connectivity or unreliable multipath reflections around corners. Under such environmental limitations, being able to leverage a momentary alignment of robots could be key to passing information, particularly when coupled with our `smart' mission management system \cite{ii2021assessment}. As a result of our partnership with Meshmerize GmBH, we were able to exploit these brief connectivity moments without pausing our mission to allow the mesh network to re-establish itself. 

\subsection{Prioritization Controls}
In testing prior to the final event, we were passing full maps (before implementing the differential maps of Section \ref{diff_maps}) from each robot to the human supervisor. As a rough estimate, transmitting full maps at 1 Hz would have required 5.4GB of mesh data. With that setup, we were quickly saturating available channel bandwidth, leading to delays in passing high-priority data such as artifact reports and telemetry. The ability to establish prioritization controls on our message traffic was a key component of our communication system's effectiveness, particularly once first-person-view was added. In our system, the prioritization was only established on a per-robot basis; there was no coordination between robots about the use of the shared medium. As a result, one robot sending FPV data could be interfering with another robot's telemetry; since our system is completely decoupled, there is no way to coordinate these types of traffic. In general, depending on the system architecture, such detailed coordination may or may not be necessary, desirable, or even feasible. However, our limited implementation in the final event has amply demonstrated that data prioritization must be a consideration in such communication-restricted environments. 

\section{Conclusions and Future Work}

In this work, we present a novel mesh networking system composed of open-source, commercial, and custom components that powered our competition team to a prize-winning victory. Our novel contribution, the \textit{udp\_mesh} transport layer, serves to link robots together by providing several communication services, including discovery, reliable transport, and prioritization, implemented over UDP. 

While \textit{udp\_mesh} performed admirably for our needs, we certainly recognize several areas in which improvements can be made. We note that fundamentally, \textit{udp\_mesh} is still reliant on a layer 2 network that is capable of carrying IP traffic; there is no specific need why this requirement must exist. Current work includes building adapters for \textit{udp\_mesh} to permit bridging multiple mesh network types, such as those implemented by low-power controllers such as ESP32 chips\footnote{\url{https://github.com/espressif/esp-mdf}}, XBee-based networks\footnote{\url{https://www.digi.com/products/embedded-systems/digi-xbee/rf-modules/2-4-ghz-rf-modules/xbee-digimesh-2-4}}, or LoRa networks \cite{cotrim2020lorawan} that do not carry IP natively. 

Particularly for use with UAVs in a heterogeneous robot team, we see mesh networking as enabling marsupial operations, in which a larger ground vehicle provides data analysis and support services for a remote UAV. With a transparent bridging capability, the UAV-to-parent link could use a completely different radio system suitable for a high-bandwidth point-to-point link while maintaining connectivity with the rest of a long-range mesh network with additional radios. 

\section*{Acknowledgments}
The contributions of the rest of the MARBLE Subterranean Challenge team were crucial in fielding our complete system and are gratefully acknowledged. 

\bibliography{main.bib}
\bibliographystyle{IEEEtran}

\end{document}

%% file: helper_figures.tex
\usepackage{comment}
\usepackage{bm}
\usepackage{outlines}
\usepackage{color}
\usepackage{hyperref}
\hypersetup{pdfborder = 0 0 0}
\usepackage{microtype} 
\usepackage{tabularx}
\usepackage{algorithm,algpseudocode}
\usepackage{adjustbox}
\usepackage[font=footnotesize]{caption}
\usepackage[rightcaption]{sidecap}
\sidecaptionvpos{figure}{t}

\usepackage{pgfplots}
\usepackage{callouts}
\usepackage{tikz}
\usepackage{pgfgantt}
\usetikzlibrary{calc, arrows,arrows.meta,positioning,graphs,automata, shapes, shapes.geometric, shapes.multipart}
\usetikzlibrary{tikzmark,arrows.meta}
\usepackage{circuitikz}
\tikzset{
    >=stealth',
    defnode/.style={
           rectangle,
           rounded corners,
           draw=black, thick,
           minimum height=2em,
           text centered},
    defedge/.style={
           ->,
           thick,
           shorten <=2pt,
           shorten >=2pt,}
}
\tikzstyle{component} = [draw, fill=blue!20, align=center, text centered, rounded corners]
\tikzstyle{human_stage} = [draw, fill=green!15, minimum width=5em, text centered, rounded corners]
\pgfdeclarelayer{background}
\pgfdeclarelayer{foreground}
\pgfsetlayers{background,main,foreground}
\RequirePackage{suffix}
\RequirePackage{soul}
\extrafloats{100}

\newcommand{\neweq}[2]{
\begin{equation}
 \label{e:#1}
 #2
\end{equation}
}

\WithSuffix\newcommand\neweq*[1]{
	$$
		#1
	$$
}

\usepackage{booktabs}
\usepackage{array}
\usepackage{colortbl}
\newcolumntype{L}[1]{>{\raggedright\let\newline\\\arraybackslash\hspace{0pt}}m{#1}}
\newcolumntype{C}[1]{>{\centering\let\newline\\\arraybackslash\hspace{0pt}}m{#1}}
\newcolumntype{R}[1]{>{\raggedleft\let\newline\\\arraybackslash\hspace{0pt}}m{#1}}
\newcolumntype{N}{@{}m{0pt}@{}}
\RequirePackage{float}
\restylefloat{table}
\newcommand{\newtable}[5][\normalsize]{
\setcounter{oli}{0}
\let\backupone\1
\let\backuptwo\2
\let\1\tablevelone
\let\2\tableveltwo
\begin{table}
	\begin{center}
		\caption{#4}
		#1
		\begin{tabular}{#3}
			#5
		\end{tabular}
		\label{t:#2}
	\end{center}
\end{table}
\let\1\backupone
\let\2\backuptwo
}

\makeatletter

\newcounter{oli}
\newcounter{olii}[oli]
\providecommand\theleveli{\@arabic\c@oli}%
\providecommand\thelevelii{\theleveli.\@arabic\c@olii}%
\makeatother

\providecommand\tablevelone{\rowcolor{gray!25}\setcounter{olii}{0}\refstepcounter{oli}\theleveli}
\providecommand\tableveltwo{\refstepcounter{olii}\thelevelii}

\RequirePackage{amsfonts}
\RequirePackage{amsmath}

\RequirePackage{graphicx}
\RequirePackage[verbose]{wrapfig}
\RequirePackage{float}
\RequirePackage{subfig}

\newcommand{\scaledfig}[4]{
	\begin{figure}[!ht]
		{\centering
			\includegraphics[width=#4\textwidth]{#1}
			\caption{#3}
			\vspace{-12pt}
			\label{f:#2}
		}
	\end{figure}
}

\newcommand{\tikzfig}[4]{
\begin{figure*}[#2]
	#4
	{\centering
	\vspace{-1em}
	\caption{#3}
	\vspace{-10pt}
	\label{f:#1}
	}
\end{figure*}
}

\newcommand{\tikzfigcol}[3]{
\begin{figure}[htb]
    {
    \centering
	#3
	\caption{#2}
	\vspace{-10pt}
	\label{f:#1}
	}
\end{figure}
}

\newcommand{\fig}[1]{\textbf{Fig. \ref{f:#1}}}

\def\fps@figure{htbp}
\def\fps@table{htbp}

\makeatletter

\newcounter{hypo}
\newcounter{subhypo}[hypo]
\makeatother

\definecolor{lgreen} {RGB}{180,210,100}
\definecolor{dblue}  {RGB}{20,66,129}
\definecolor{ddblue} {RGB}{11,36,69}
\definecolor{lred}   {RGB}{220,0,0}
\definecolor{nred}   {RGB}{224,0,0}
\definecolor{norange}{RGB}{230,120,20}
\definecolor{nyellow}{RGB}{255,221,0}
\definecolor{ngreen} {RGB}{98,158,31}
\definecolor{dgreen} {RGB}{78,138,21}
\definecolor{nblue}  {RGB}{28,130,185}
\definecolor{jblue}  {RGB}{20,50,100}

\definecolor{GreenYellow}       {RGB}{217, 229, 6} 	    
\definecolor{Yellow}            {RGB}{254, 223, 0} 	    
\definecolor{Goldenrod}         {RGB}{249, 214, 22} 	
\definecolor{Dandelion}         {RGB}{253, 200, 47} 	
\definecolor{Apricot}           {RGB}{255, 170, 123} 	
\definecolor{Peach}             {RGB}{255, 127, 69} 	
\definecolor{Melon}             {RGB}{255, 129, 141} 	
\definecolor{YellowOrange}      {RGB}{240, 171, 0} 	    
\definecolor{Orange}            {RGB}{255, 88, 0} 	    
\definecolor{BurntOrange}       {RGB}{199, 98, 43} 	    
\definecolor{Bittersweet}       {RGB}{189, 79, 25} 	    
\definecolor{RedOrange}         {RGB}{222, 56, 49} 	    
\definecolor{Mahogany}          {RGB}{152, 50, 34} 	    
\definecolor{Maroon}            {RGB}{152, 30, 50} 	    
\definecolor{BrickRed}          {RGB}{170, 39, 47} 	    
\definecolor{Red}               {RGB}{255, 0, 0}        
\definecolor{BrilliantRed}      {RGB}{237, 41, 57} 	    
\definecolor{OrangeRed}         {RGB}{231, 58, 0} 	    
\definecolor{RubineRed}         {RGB}{202, 0, 93}       
\definecolor{WildStrawberry}    {RGB}{203, 0, 68} 	    
\definecolor{Salmon}            {RGB}{250, 147, 171} 	
\definecolor{CarnationPink}     {RGB}{226, 110, 178} 	
\definecolor{Magenta}           {RGB}{255, 0, 144} 	    
\definecolor{VioletRed}         {RGB}{215, 31, 133} 	
\definecolor{Rhodamine}         {RGB}{224, 17, 157} 	
\definecolor{Mulberry}          {RGB}{163, 26, 126} 	
\definecolor{RedViolet}         {RGB}{161, 0, 107} 	    
\definecolor{Fuchsia}           {RGB}{155, 24, 137} 	
\definecolor{Lavender}          {RGB}{240, 146, 205} 	
\definecolor{Thistle}           {RGB}{222, 129, 211} 	
\definecolor{Orchid}            {RGB}{201, 102, 205} 	
\definecolor{DarkOrchid}        {RGB}{153, 50, 204} 	
\definecolor{Purple}            {RGB}{182, 52, 187} 	
\definecolor{Plum}              {RGB}{79, 50, 76} 	    
\definecolor{Violet}            {RGB}{75, 8, 161} 	    
\definecolor{RoyalPurple}       {RGB}{82, 35, 152} 	    
\definecolor{BlueViolet}        {RGB}{33, 7, 106} 	    
\definecolor{Periwinkle}        {RGB}{136, 132, 213} 	
\definecolor{CadetBlue}	  	    {RGB}{95, 158, 160} 	
\definecolor{CornflowerBlue}  	{RGB}{99, 177, 229} 	
\definecolor{MidnightBlue}	  	{RGB}{0, 65, 101} 	    
\definecolor{NavyBlue}          {RGB}{0, 70, 173}       
\definecolor{RoyalBlue}         {RGB}{0, 35, 102}       
\definecolor{Blue}              {RGB}{0, 24, 168}       
\definecolor{Cerulean}          {RGB}{0, 122, 201}      
\definecolor{Cyan}              {RGB}{0, 159, 218}      
\definecolor{ProcessBlue}       {RGB}{0, 136, 206}      
\definecolor{SkyBlue}           {RGB}{91, 198, 232}     

\definecolor{Turquoise}         {RGB}{0, 255, 239} 	    

\definecolor{TealBlue}          {RGB}{0, 124, 146} 	    
\definecolor{Aquamarine}        {RGB}{0, 148, 179} 	    
\definecolor{BlueGreen}         {RGB}{0, 154, 166} 	    
\definecolor{Emerald}           {RGB}{80, 200, 120} 	
\definecolor{JungleGreen}       {RGB}{0, 115, 99} 	    
\definecolor{SeaGreen}          {RGB}{0, 176, 146} 	    
\definecolor{Green}             {RGB}{0, 173, 131} 	    
\definecolor{ForestGreen}       {RGB}{0, 105, 60} 	    
\definecolor{PineGreen}         {RGB}{0, 98, 101} 	    
\definecolor{LimeGreen}         {RGB}{50, 205, 50} 	    
\definecolor{YellowGreen}       {RGB}{146, 212, 0} 	    
\definecolor{SpringGreen}       {RGB}{201, 221, 3} 	    
\definecolor{OliveGreen}        {RGB}{135, 136, 0} 	    
\definecolor{RawSienna}         {RGB}{149, 82, 20} 	    
\definecolor{Sepia}             {RGB}{98, 60, 27} 	    
\definecolor{Brown}             {RGB}{134, 67, 30}      
\definecolor{Tan}               {RGB}{210, 180, 140}	
\definecolor{Gray}              {RGB}{139, 141, 142} 	
\definecolor{DarkSlateGray}     {RGB}{57, 87, 86} 	    

\definecolor{CeruleanFrost}     {RGB}{106, 159, 200}     
\definecolor{PastelPink}        {RGB}{211, 162, 156}     

\definecolor{Black}		  	    {RGB}{30, 30, 30}       
\definecolor{White}		  	    {RGB}{255, 255, 255}    